\pdfoutput=1
\documentclass[twocolumn,aps]{revtex4}

\usepackage{graphicx}
\usepackage{amsmath}
\usepackage{mathrsfs}
\usepackage{chapterbib}

\begin{document}

\title{Phonon cooling of nanomechanical beams with tunnel junctions}

\author{P.~J.~Koppinen}
\author{I.~J.~Maasilta}
\affiliation{
Nanoscience Center, Department of Physics, University of Jyv\"askyl\"a, P.~O.~Box 35, 
FIN--40014 University of Jyv\"askyl\"a, Finland.
}

\date{\today}
\begin{abstract}
We demonstrate electronic cooling of 1D phonon modes in suspended nanowires for the first time, using normal metal--insulator--superconductor (NIS) tunnel junctions. Simultaneous cooling of both electrons and phonons to a common temperature was achieved. In comparison with non-suspended devices, better cooling performance is achieved in the whole operating range of bath temperatures between 0.1-0.7 K.  The observed low-temperature thermal transport characteristics are consistent with scattering of ballistic phonons at the nanowire-bulk contact as being the mechanism limiting thermal transport. 
At the lowest bath temperature of the experiment $\sim$ 100 mK, both phonons and electrons in the beam were cooled down to 42 mK, which is below the refrigerator base temperature.       
\end{abstract}

\maketitle
 
The ability to cool both the electrons and phonons of a mesoscopic device below the temperature of its surrounding bath is potentially valuable for all areas of low-temperature physics, and could prove especially useful for applications such as  nanobolometry in sub-millimeter telescopes \cite{wei,kenyon}, quantum computing \cite{review} and cooling of nanomechanical oscillators into their quantum mechanical ground state \cite{roukesphystoday}. In the sub-Kelvin temperature range, the thermal coupling between a nanoscale device and its thermal bath becomes very weak \cite{schwabphonons,karvonenPRL,wei}, leading easily to overheating problems due to even tiny dissipated power levels ($\sim$ fW)  from the measurement signals and external noise. A cooling method, where the cooling power is applied directly to the device can therefore be extremely valuable, as it actually takes advantage of the weakness of the thermal coupling.    

A tunnel junction can be used for direct cooling at sub-Kelvin temperatures, if one of the electrodes is superconducting and the other in normal state, i.e. with a normal metal--insulator--superconductor (NIS) structure \cite{martiniscooler,leivocooler,jukkareview} (Fig. 1). If fabricated on bulk substrates, only sizeable electron cooling can be achieved due to the strong weakening of electron--phonon interaction with temperature \cite{roukesep,karvonenPRL,courtphon}.  
To be able to cool the phonons, the phonon thermal conductance out of the device needs to be small enough to become the  
bottleneck for the heat flow. Using this idea, cooling of phonons in thin but large insulating membranes has been demonstrated with large area tunnel junction coolers \cite{luukanenmembrane,ullommembrane,ullomnew}, but direct phononic cooling of a nanoscale (1D) device has only been suggested theoretically \cite{hekkjustatheory}. In this work, we demonstrate significant NIS junction cooling of both electrons and phonons in suspended 1D nanowires to a common temperature below the bath temperature of the refrigerator, and present quantitative results on the phonon thermal transport processes involved (ballistic phonon transmission). In our 1D geometry, phonon cooling is approximately two orders of magnitude more effective than in the 2D case \cite{ullomnew} in terms of the cooling power required to achieve a given temperature reduction.     
\begin{figure*}[!htp]
\includegraphics[height=9.25cm]{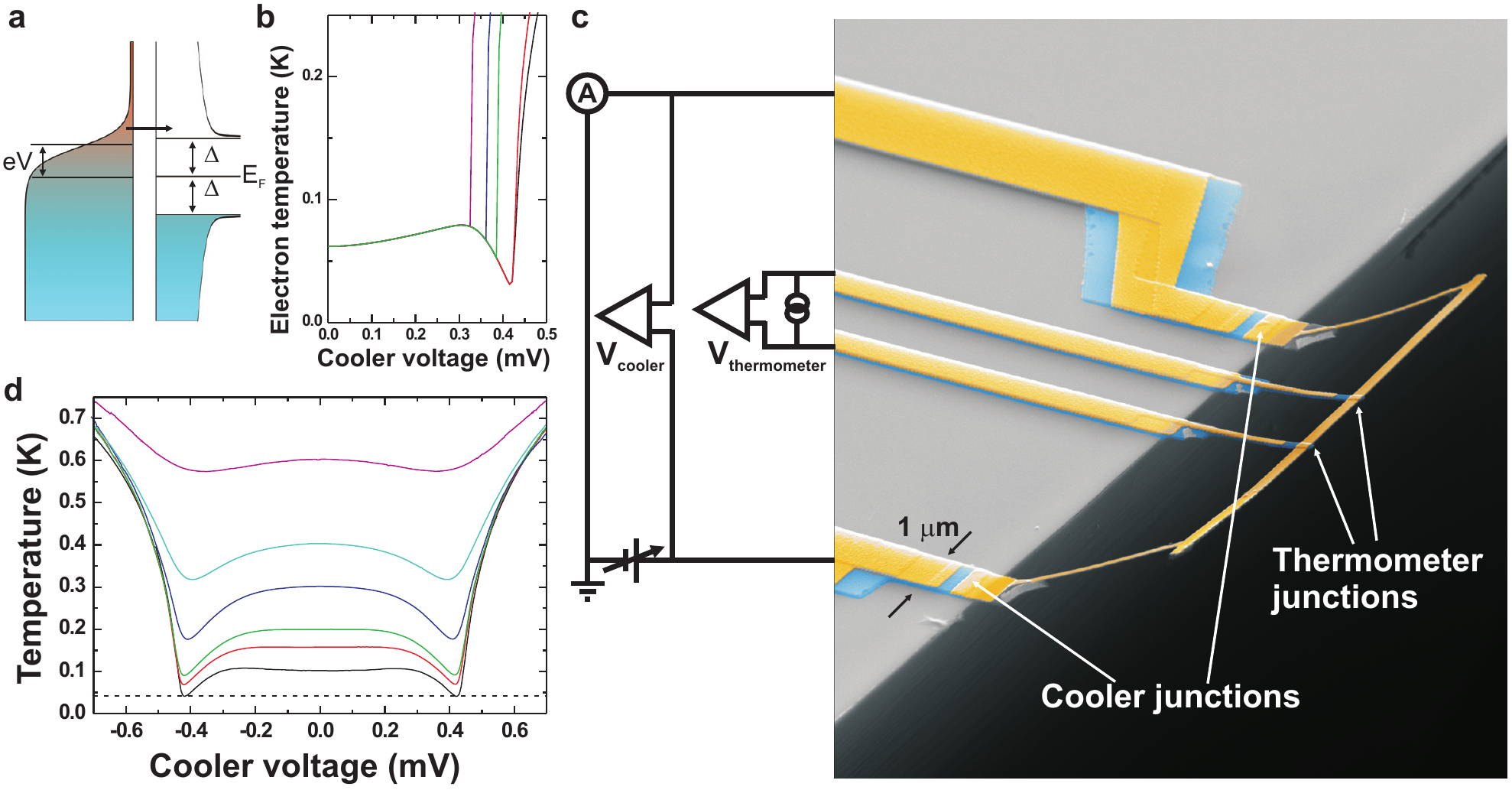}
\caption{\label{fig1} (Color online)(a) Schematic of the operation principle of an NIS tunnel junction cooler. The superconducting gap $\Delta$ filters through only hot electrons from the normal metal. Heat is removed from the normal metal and injected into the superconductor, which must be efficiently removed by quasiparticle traps \cite{jukkareview}. 
(b) Theoretically calculated cooling curves vs. bias voltage with different hot quasiparticle removal (trapping) efficiencies. The curves with less cooling correspond to worse trapping efficiency (effectively hotter superconductor).  (c) Schematic of the measurement setup and a scanning electron micrograph of a typical suspended sample.  
Yellow color indicates Cu (normal metal), blue Al (superconductor). Cooler junctions are located at the edge of the substrate, whereas the thermometer junctions are at the center of the nanowire. The Cu on top of the Al leads serves as quasiparticle trap 
for both the cooler and thermometer junctions. 
(d) Measured temperature of a typical nanowire sample as a function of cooler voltage at bath temperatures 50,160,200,300,400 and 600 mK, 
dashed line corresponds to 42 mK.}
\end{figure*}

All suspended nanowires have a length of either 10 or 20 $\mu$m, width 200 or 300 nm and thickness 60 nm, and were  
fabricated to form bilayers of evaporated copper (30 nm) on silicon nitride (30 nm) using e-beam lithography, vacuum evaporation and reactive ion etching (see \cite{supplement}). The  Cu/SiN wire is connected to the substrate by four free--standing bridges with a length 5 $\mu$m, and width 150 nm  (Fig. \ref{fig1} (c)). The outer bridges are also Cu/SiN bilayers of thickness 60 nm, and connect the Cu wire (total length 20 or 30 $\mu$m) to wider (1 $\mu$m) superconducting Al electrodes on the bulk substrate via two larger area (0.35 ($\mu$m$)^2$) NIS--junction coolers. The cooler junctions must be located on the bulk substrate to avoid serious back-flow of dissipated heat from the superconductor into the nanowire\cite{LT25}. The inner bridges have a composition Cu/Al/SiN (thickness 90 nm), connecting Al leads with Cu quasiparticle traps to two smaller (0.05 ($\mu$m)$^2$) thermometer NIS--junctions located on the suspended nanowire.  Since the measured phonon thermal conductance from the narrow bridges to the bulk substrate is approximately an order of magnitude smaller than the thermal conductance between the electrons and the phonons (detailed discussion below), phonon transmission becomes the thermal bottleneck. Thus, both the electrons and the phonons in the suspended wire have a common temperature and can be cooled simultaneously in our sample geometry, in contrast to a recent report on electron cooling in nanowires \cite{muhonen}.  

Measurements were perfomed in a $^3$He--$^4$He dilution refrigerator with a base temperature of $\sim$ 50 mK with several  
stages of filtering in the wires\cite{supplement}. In the experiment, the temperature of the nanowire was measured as a function of the bias voltage across the cooler junctions\cite{supplement}. Schematic of the measurement setup is shown in Fig. 
\ref{fig1} (c). Using this simple measurement, we obtain temperature response curves that typically look like Figure \ref{fig1} 
(d) as a function of the cooler bias and refrigerator bath temperature $T_{bath}$.  As expected from theory \cite{jukkareview}, the maximum 
cooling is obtained at cooler voltage $V \sim 2\Delta/e$, where $\Delta = 215$ $\mu$eV for our samples. Two main observations are immediately apparent: (i) At the lowest bath temperature 50 mK, where the nanowire has a temperature 100 mK at zero cooler bias (due to noise power of $\sim 6$ fW radiated from higher temperature parts of the circuit), the lowest temperature achieved around $V \sim 2\Delta/e$ was 42 mK, a reduction of $\sim 60$ mK. (ii) The cooler clearly still works at bath temperatures as high as 600 mK. These are the main results of this work, and in the following we elaborate on the physics by comparing samples of different sizes and results between suspended wires and wires on bulk substrates.    

\begin{figure}
\includegraphics[width=8cm]{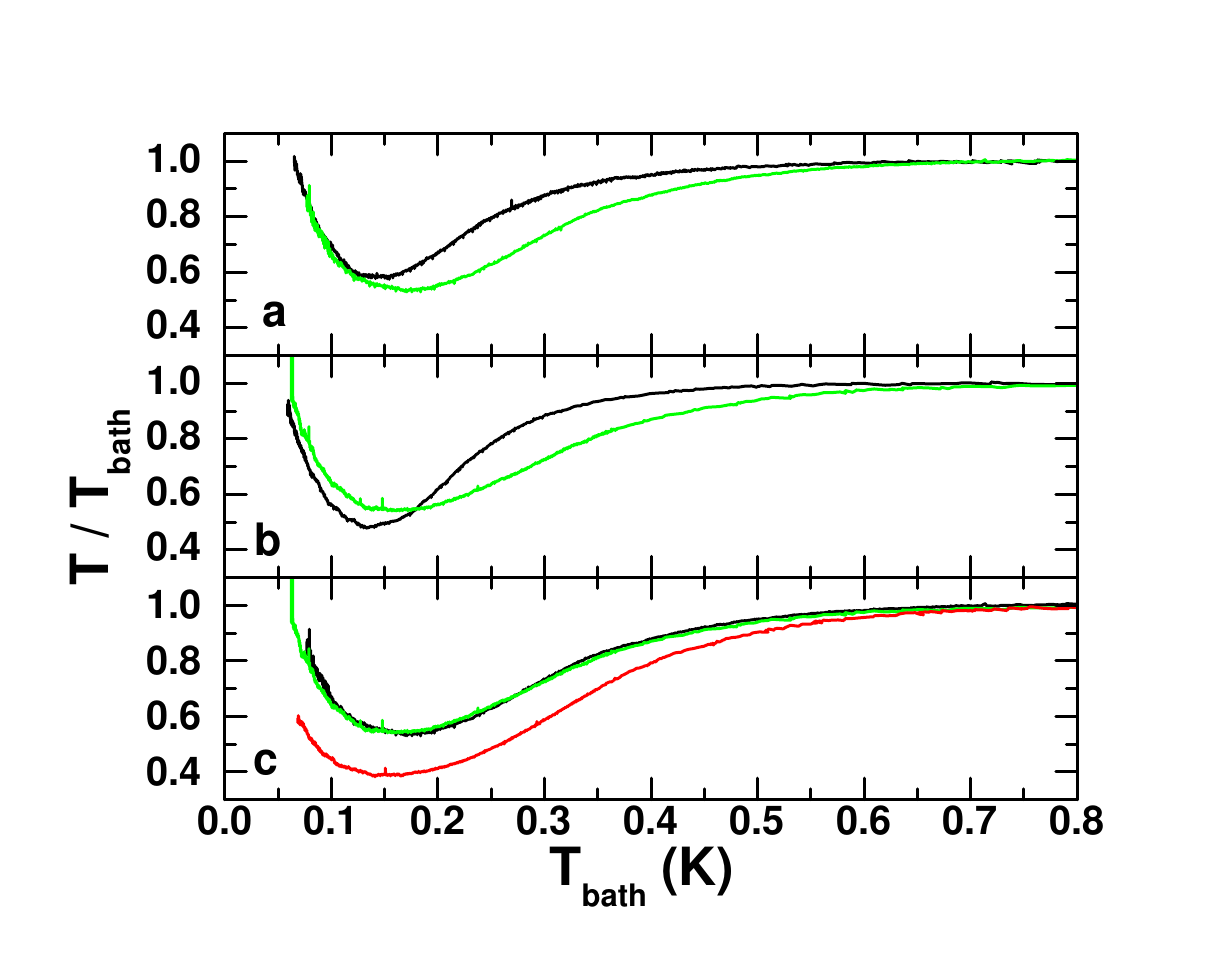}
\caption{\label{fig2} (Color online)(a) Temperature of a cooled suspended (green) and bulk  
(black) 20 $\mu$m long nanowires normalized with the bath temperature $T/T_{bath}$ vs. $T_{bath}$. The cooler is biased to the 
optimal cooling voltage, while the bath temperature is changed. Both samples have an electron gas volume $\Omega=0.17 \mu$m$^3$ 
and a tunneling resistances $R_T \sim 3$ k$\Omega$ for suspended and $R_T \sim 4.4$ k$\Omega$ for bulk sample. At the low temperature regime, cooling efficiency is limited by the junctions, whereas at higher temperatures cooling behavior is different due to the different dissipation mechanisms (e-p 
interaction for bulk samples vs. phonon transport for suspended samples). (b) Same, but for longer 30 $\mu$m long wires with 
$\Omega=0.36$ $\mu$m$^3$ and $R_T\sim 3$ k$\Omega$.  (c) Same, comparing the two suspended samples in (a) (black) and (b) (green) with a third suspended sample (red) with $\Omega=0.36$ $\mu$m$^3$  and $R_T$=1.7 k$\Omega$. 
}
\end{figure}

Cooler samples on bulk substrates were fabricated with the same metal film and tunnel junction geometry as the suspended samples during the same fabrication run \cite{supplement}. Figure \ref{fig2} (a) shows the measured cooling (ratio of measured temperature to the bath temperature) vs. bath temperature at optimal cooling bias for a suspended (green) and a bulk (black) sample, both with wire lengths 20 $\mu$m and similar junction properties,  whereas  Fig. \ref{fig2} (b) shows it for  longer 30 $\mu$m wires. Clearly, the suspended samples in both cases show better cooling ($\sim 20 $ \% improvement at 300 mK) extending to much higher bath temperatures (up to $\sim$ 600 mK), pointing to a difference in the dissipation mechanisms between the bulk and suspended samples. A further confirmation of the differences can be seen by comparing the $T(V_{cooler})$ curves \cite{supplement}. In addition,  in Fig. \ref{fig2} (c) we compare cooling for three suspended samples in pairs: Two samples have different nanowire sizes but the same tunnel junction properties ($L=$20 $\mu$m, $w=$200 nm (black) and $L=$30 $\mu$m, $w=$ 300 nm (green), $R_T \sim$3 k$\Omega$), whereas the third sample (red) has the same size as the longer wire of the previous pair, but more transparent tunnel barriers ($R_T \sim$1.7 k$\Omega$). It is quite clear that the size of the suspended nanowire has no effect on the cooling behavior, whereas the tunnel resistance $R_T$ has a stronger effect, as expected by theory \cite{supplement}. This size--independence proves that neither 3D \cite{roukesep} nor 1D \cite{hekkjustatheory} electron--phonon (e-p) interaction limits the heat flow in the suspended samples, unlike in bulk coolers, where a strong volume dependence is observed \cite{jukkareview} due to the 3D e-p interaction (clearly noticeable by comparing the bulk results in Figs \ref{fig2} (a) and (b)). We can thus deduce that electrons and phonons in the suspended wire are in quasiequilibrium  (common temperature) and are therefore both cooled simultaneously. Moreover, we have  observed that there are no  temperature gradients within the wire  \cite{supplement} so that electronic diffusion \cite{jltp} is not operational either.  This means that heat flow is limited by phonon transmission (phonon thermal conductance), and must be dominated by the interface between the suspended wire and the bulk because of the lack of wire length dependence.  

In quantitative terms, the heat flow between the substrate and the suspended nanowire is determined by the power balance  
condition $P_{heat}=P_{cool}$, where the heat flow from the surroundings $P_{heat}$ equals the cooling power of the junctions  
$P_{cool}=2\dot {Q}_{cool}$, where $\dot {Q}_{cool}$ is the cooling power of a NIS junction\cite{supplement}.    
Regardless of the limiting heat transport mechanism (e-p interaction or phonon thermal conductance), we can generally write 
\begin{equation}\label{model}
2\dot {Q}_{cool} = A(T_{bath}^{n}-T_{nw}^{n})+\beta(2\dot{Q}_{cool}+IV_{cool}),
\end{equation}
where the first term on the right describes the the heating from the environment ($T_{nw}$ is the nanowire temperature), and the second backflow of dissipated heat $2\dot{Q}_{cool}+IV_{cool}$ from the superconductor to the normal metal (0 $\le \beta \le$ 1) due to non-equilibrium recombination phonons and back-tunneling \cite{ullommembrane2}. Here $I$ is the current though the cooler junctions and $V_{cool}$ the voltage across them, $A$ is a parameter describing the strength of the heat flow between the nanowire and the bath, and the exponent $n$ depends on the heat flow mechanism. In the calculation of $\dot{Q}_{cool}$ we have taken into account a measured broadening of the quasiparticle density of states due to lifetime-effects \cite{Dynes} and/or two-electron Andreev processes \cite{hekking} \cite{supplement}, because of its strong influence on cooling efficiency \cite{jukkareview,courtois}. In the case of 3D (1D) electron--phonon mediated heat flow  $A=\Sigma\Omega$, where $\Sigma$ is the electron--phonon coupling constant and $\Omega$ the electron gas volume (length for 1D), and  $n=3-6$ depending on the level of purity of the metal film and the phonon dimensionality \cite{eppower,karvonenPRL,hekkjustatheory}. In contrast, if the heat flow is limited by phonon transmission at the nanowire-bulk boundary (suspended samples), $A$ is not dependent on $\Omega$, and $n=2-6$ depending on phonon mode and dimensionality \cite{schwabphonons,kuhn2,cross,chang}. 

\begin{figure}
\includegraphics[width=8cm]{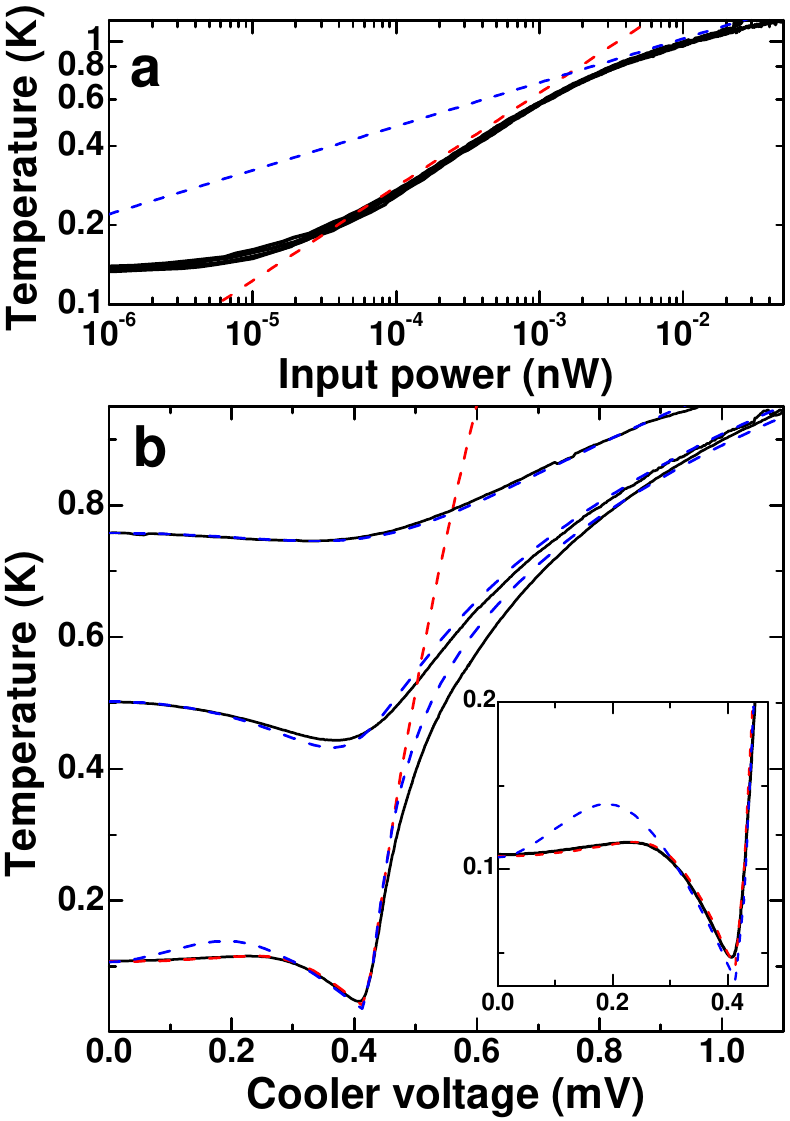}
\caption{\label{fig5} (Color online) (a) Heating experiment: temperature of a 
suspended nanowire as a function of input DC power. Red and blue dashed lines correspond power laws $n=$2.8 and $n=6$, 
respectively. The low temperature saturation is likely due to absorbed noise power of $\sim 6$ fW. $A=4.3$ 
pW/K$^{2.8}$ for the low-temperature power law $n=2.8$. (b) Temperature of a suspended nanowire as a function of cooler voltage, experimental data is shown as  black solid line. The red and blue dashed lines correspond to thermal model of eq. \ref{model} with $n=$2.8 and $n=6$, respectively. For the $n=2.8$ model, we used the value for $A$ determined from the heating experiment (no fitting). Inset is zoom--in to the cooling region of the low-temperature data in (b).}
\end{figure}

To understand the transport process,  we performed a complementary  
heating experiment without cooler tunnel junctions but with the same nanowire width $w= 300$ nm and thickness $t=60$ nm (length 
was $L=$ 24 $\mu$m). In that experiment we substituted the cooler tunnel junctions by direct contact between the normal metal (Cu) and a superconductor (Nb). These NS--junctions work as good electrical contacts but, on the other hand, as nearly perfect thermal barriers so that Joule heating power $P_{in}=IV$ is dissipated uniformly in the normal metal and $P_{in}=A  
(T_{nw}^n-T_{bath}^n)$. (details in \cite{supplement}). Fig. \ref{fig5} (a) shows the result of such an experiment, where the  
temperature of the nanowire is plotted as a function of the heating power in log-log scale. We notice that the data is well  
described by a transition from a power law with $n =$ 2.8 at low temperatures to a power law with  $n \sim$ 6 at high  
temperatures. The thermal conductance $G=dP/dT$ at 200 mK is 0.4 pW/K (at 100 mK extrapolated to 0.12 pW/K), which can be compared with a value for calculated 1D e-p limited conductance \cite{hekkjustatheory} (using known Cu parameters) at 200 mK, 5.2 pW/K (at 100 mK 1.3 pW/K), confirming that phonon transport limits the heat flow at low temperatures. 
 
The measured power laws for suspended nanowires can then be used in the thermal model of the coolers, Eq \ref{model}. Figs.  
\ref{fig5} (b) and (c) show cooler data in comparison with the two power laws, $n=2.8$ (red) and $n=6$ (blue). Good agreement is achieved with the low bath temperature data, if a transition from a $n=2.8$ power law into a $n \sim 6$ is assumed, in agreement with the heating experiment. We would like to stress that simply fitting the cooler data with different power laws directly is nearly impossible, as the cooler model is quite insensitive to the value of $n$. At bath temperatures $T_{bath}> 0.4 K$, $n=6$ fits fairly well the full range of bias values, but is not consistent with an e-p limited heat flow, as the parameter $A$ is much smaller than what is expected from e-p theory \cite{supplement}.  

To understand this behavior, we note that the phonons in the wire have a cross-over from 3D to 1D behavior when the thermal  
wavelength of the lowest energy transverse modes $\lambda_T=hc_t/(2.8k_BT)$ becomes larger than the wire thickness and width,  
which is estimated to take place around $T \sim 0.4$ K  using a value $c_t= 4300$ m/s for the Cu/SiN bilayer. This estimate is not far from the observed transition temperature seen in Fig. \ref{fig5} (b). Thus, we believe that in the low temperature regime $T < 0.4$ K, the nanowire phonons are one-dimensional. In the ballistic 1D limit with no scattering at the nanowire-bulk contact $n=2$ is expected \cite{schwabphonons}. However, in our sample geometry the nanowire-bulk contact is abrupt, leading to strong scattering and predicted power laws between $n=2.5-3.5$ for 1D-2D scattering \cite{cross}, or $n=4-6$ for 1D-3D scattering \cite{chang}. We thus conclude that our observed $n =2.8$ is consistent with 1D-2D boundary limited phonon scattering, which is plausible based on the device details \cite{supplement}. The value of the measured thermal conductance (Fig. \ref{fig5} (a)) per conduction channel $G/16$ (four legs with four phonon branches) is also consistent with boundary scattering, as it can be expressed in units of the quantum of thermal conductance \cite{schwabphonons} $G_0=\pi^2k_B^2T/(3h)$ as $G/16 \sim 0.13 G_0$ at 200 mK.   

In conclusion, we have demonstrated electronic cooling of phonon modes of a suspended nanowire,  where scattering of phonons at the nanowire-bulk contact limits the thermal transport. In comparison with non-suspended devices, better cooling performance is achieved in the whole operating range of bath temperatures. The 
minimum temperature reached in our best device was 42 mK, starting from an initial temperature of 100 mK. 
This limit is mostly determined by the superconducting material itself (Al), and thus the minimum temperature could be extended below 10 mK by using an additional tunnel junction cooler with a lower superconducting gap \cite{note}.    The advantages of tunnel junction coolers are:   
(i) ease of integration into a nanoscale system and compatibility with existing fabrication processes, (ii) simple operation by only DC voltage source, and (iii) the ability to cool both electrons and phonons simultaneously. 
This last point is significant for ultrasensitive nanobolometers \cite{wei,kenyon}, as both the electron and the phonon temperatures contribute to their performance (noise equivalent power), and also for nanomechanical oscillators, where the phonon modes need to be cooled.

We thank Herve Courtois, Frank Hekking, Thomas K\"uhn and Jukka Pekola for useful discussions. This work was supported by the  
Academy of Finland projects No. 118665 and 118231.


\begin{widetext}
\newpage
\end{widetext}
\setcounter{figure}{0}
\begin{center}
{\bf \large Supplementary Data for the paper "Phonon cooling of nanomechanical beams with tunnel junctions"
(EPAPS No. E-PRLTAO-102-035919)}
\end{center}

\section{Device fabrication}
The suspended samples were fabricated on single crystal (001) Si wafers with double-sided 30 nm thick low-stress silicon nitride (SiN) films grown by low--pressure chemical vapour deposition (LPCVD) and purchased from the Microfabrication Laboratory at UC Berkeley. A $500 \times 500 \mu$m square was first patterned and etched onto the bottom-side nitride, using  photolithography and reactive ion etching in CHF$_3$ gas. The remaining SiN then served as an etch mask for a  crystallographic potassium hydroxide (KOH) wet etch through the whole wafer thickness, producing  a 30 nm thick suspended SiN membrane of size $50 \times 50 \mu$m on the front side (bulk micromachining). The metallic wires were then deposited on the membrane, using electron beam lithography and two-angle shadow mask technique, where the the two metals aluminium (Al) and copper (Cu) were e-beam evaporated from different angles in an UHV system with a base pressure $10^{-8}$ mbar.  Al was evaporated first (0.2 nm/s) at 60$^{\circ}$ angle with respect to the normal of the substrate, after which it was thermally oxidized in 10 mbar for 4 min. Then Cu was evaporated with rate 0.15 nm/s from 0$^{\circ}$ angle. This produced (after lift-off) the narrow nanowire, the smaller Al/AlOx/Cu thermometer junctions (typical $R_T \sim 30-50$ k$\Omega$) in the middle of the wire, and the larger Al/AlOx/Cu cooler junctions (typical $R_T \sim 1.5-3$ k$\Omega$) at the ends of the wire (Fig. 1(c), main text). After metallization, the suspended structure was released by reactive ion etching the SiN in a CHF$_3$ plasma ( 0.1 mbar for 100 s), where the metal wires themselves served as the etching mask. Note that this process also etches down the SiN layer on the bulk substrate area around the wider leads. 

The bulk samples were fabricated on the same 30 nm LPCVD-nitridized Si wafers during the same evaporation runs, but without any 
etching processes. 

\begin{figure}[ht!]
\includegraphics[width=8.6cm]{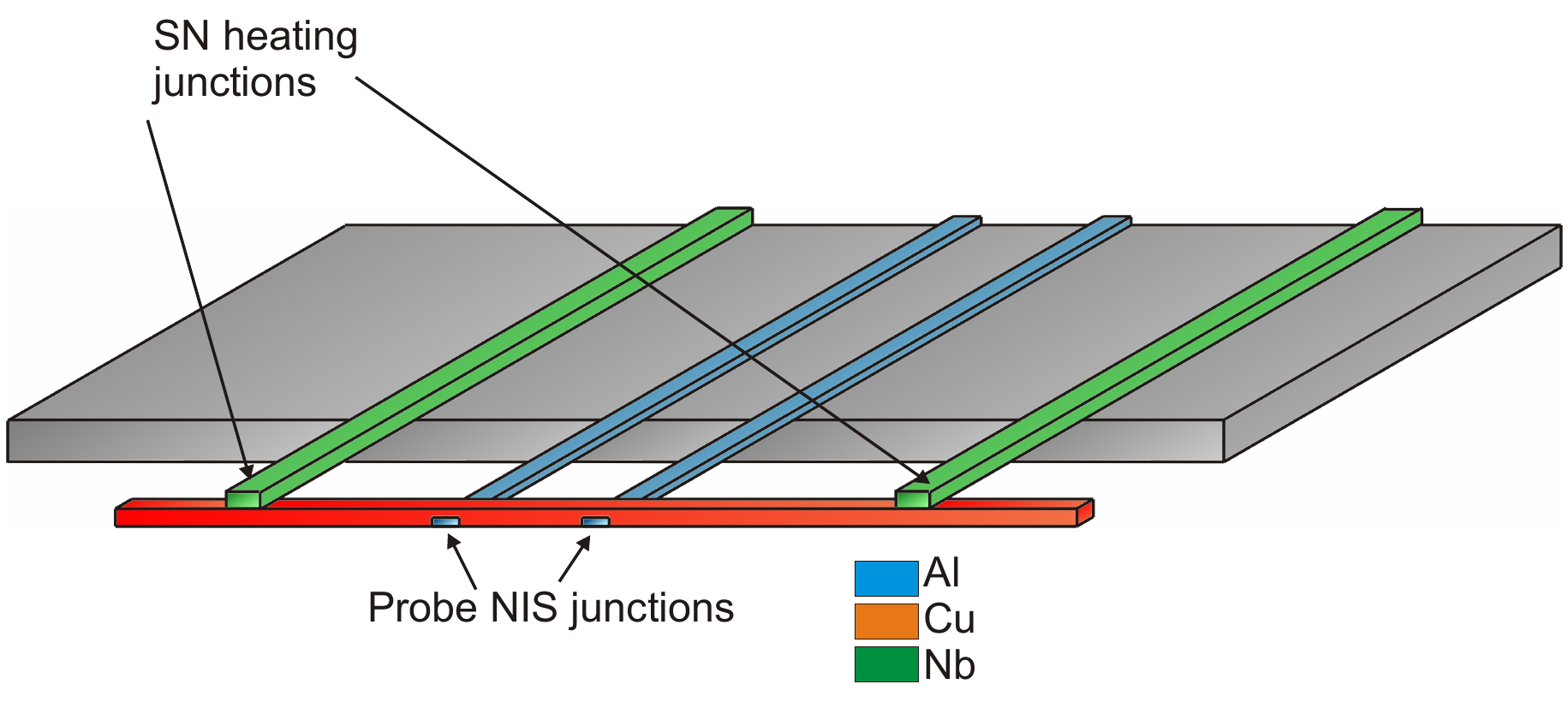}
\caption{\label{schemaSN}A Schematic of the sample with two SN heating junctions.}
\end{figure}

Samples used in the heating experiment containing direct superconductor--normal (SN) junctions had a more complex fabrication 
procedure than the coolers, mainly because three different materials were used: Cu as the normal metal, Al as the superconductor for NIS thermometers and Nb as the superconductor for the heating probes connecting directly to Cu. The thermometer junctions were still the same Al--AlO$_{x}$--Cu SIN junctions as in the coolers, and thus the first evaporation and oxidation steps of Al were the same as for the cooler samples, (60$^{\circ}$ angle, thermal oxidation in 10 mbar for 4 min).  Then the sample stage was rotated horizontally by 90$^{\circ}$ and Cu deposited from tilt--angle of 60 $^{\circ}$. After that, the sample was again rotated horizontally by 45$^{\circ}$ and deposition of 30 nm of Nb with rate of 0.5 nm/s followed from the tilt--angle of 60$^{\circ}$. The structural release process was the same as for the suspended coolers. A Schematic view of the resulting heating sample is shown in Fig. \ref{schemaSN}. Nb was used for the heating junctions because of its high superconducting gap, which prevents any heat from leaking into the superconductor due to multiple Andreev reflections \cite{SHoffmann}. This means that the dissipated power within the normal metal nanowire can be accurately determined by simply measuring the I--V characteristics of the SNS structure, and calculating the dissipated power as $P_{\mathrm{diss}}=IV$.

\section{Operation of the Cooler}

The basic principle of cooling of a NIS tunnel junction is based on the existence of the superconducting energy gap $\Delta$ for single particle electronic excitations (Fig. 1(a), main text). An electron from the normal metal cannot enter the superconductor,  
unless it has at least energy $\Delta$. A voltage bias $V$ can supply this energy so that at $T=0$ current can flow if $eV >  
\Delta$. However, at finite temperatures electrons follow the Fermi-Dirac distribution, so that even at biases $eV < \Delta$ there are some energetic (hot) electrons that can tunnel. As only hot electrons escape in that case, the temperature of the remaining electrons in the normal metal is lowered, and the magnitude of bias dependent heat flow (power) due to single particle tunneling from the normal metal to the superconductor is \cite{SJukkareview}
\begin{eqnarray}\label{Pcool}
\lefteqn{\dot{Q}(V)=\frac{1}{e^2R_{T}}\int^{\infty}_{-\infty}(E-eV)g_{S}(E)} \nonumber \\ 
& &[f_{N}(E-eV,T_{N})-f_{S}(E,T_{S})]dE, 
\end{eqnarray} 
where $R_{T}$ is the tunneling resistance of the junction, $g_{S}(E)$ the density of states (DOS) in the superconductor and  
$f_{N,S}$ the Fermi distributions of normal metal and superconductor, respectively. Eq. (2) is valid as long as both the normal 
metal and the superconductor remain in quasiequilibrium, i.e. as long as the tunneling rate is not much larger than the  
electron-electron scattering rate \cite{SJukkareview}. The optimal cooling power is achieved with a bias close to the edge of the superconducting gap $V \sim \Delta/e$ , and by increasing the bias further heating starts to dominate (Fig. 1(b), main text). It is also clear that to maximize the cooling power, the superconducting electrode temperature $T_S$ has to be as low as possible, which can be ensured by  removing the excess injected quasiparticles as efficiently as possible, for example, by letting them quickly diffuse out into another normal electrode (quasiparticle trap) \cite{SJukkareview}. For practical purposes, it is useful to connect two NIS junctions in series (SINIS), since the cooling power is doubled in that case.
 
The cooling curves $T(V)$ (Fig. 1(d), main text) were measured by slowly sweeping  the voltage bias of the two cooler junctions in series (SINIS) $V_{\mathrm{cooler}}$ and  measuring the resulting temperature with the second, smaller pair of NIS junctions (SINIS thermometer). This was performed at different refrigerator bath temperatures  $T_{\mathrm{bath}}$. In addition, we also performed measurements where the cooler voltage bias $V_{\mathrm{cooler}}$ was kept constant at optimal cooling point while changing $T_{\mathrm{bath}}$ slowly and measuring the temperature response (Fig. 2, main text). This way, the performance as a function of $T_{\mathrm{bath}}$ can be determined more accurately. $V_{\mathrm{cooler}}$ and $V_{\mathrm{therm}}$ were both measured with a high input impedance differential voltage preamplifier (Ithaco 1201). Measurements of the $I(V)$ characteristics of the cooler junctions also show consistent cooling behavior in comparison with the separate temperature measurement. 

Because of the extreme sensitivity of the samples even to tiny heating powers $\sim 10$ fW, extra care was taken to filter out  
unwanted external RF noise from the experiment. The setup uses shielded coax wiring from room temperature to 4K flange, where a 
$RC$-filter stage is located. From 4K to 50 mK, the wires are Thermocoax cables of length $\sim $1.5 m, with its known good attenuation properties at high frequencies. In addition, another set of RC filters are located at the 50 mK sample stage, inside a copper box thermalized to $T_{\mathrm{bath}}$. 

\section{Comparison of the cooling curves of bulk and suspended coolers}

\begin{figure}
\includegraphics[width=8.6cm]{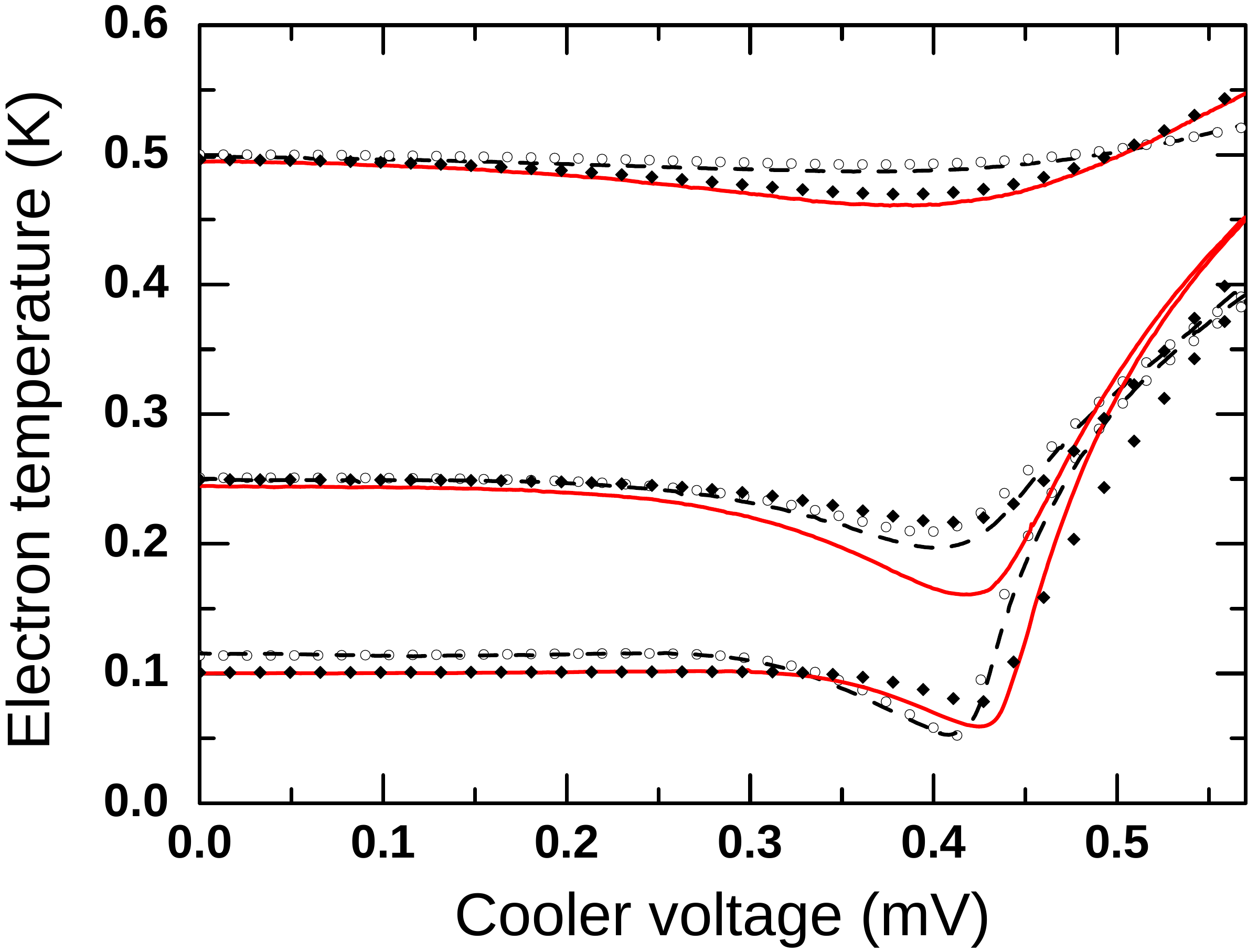}
\caption{\label{fig4} 
Cooler sweeps ($T$ vs. $V$) for a bulk (black, dashed)  
and a suspended (red, solid) sample with $R_T \sim 3$k$\Omega$. Open circles were calculated from the thermal model  (Eq. 1 main text) using typical 3D electron--phonon coupling parameters for Cu  $n$=5 and $\Sigma=2.1 \cdot 10^9$ WK$^{-5}$m$^{-3}$. The fitted value for $\beta=0.03$ Diamonds are the 1D case, with no additional adjustable parameters.} 
\end{figure}

The main text discussed the differences between bulk and suspended samples in terms of measurements, where the cooler bias was 
kept at optimum value, and bath temperature was varied (Fig. 2, main text). 
A further confirmation of the difference between bulk and suspended samples is shown in  Fig. \ref{fig4}, where measured cooling curves ($T$ vs. $V$) for the bulk (dashes, black line) and the suspended (solid, red line) sample of similar values of $R_T$ are plotted for three different bath temperatures, and compared with the thermal model (Eq. 1 of main text) limited either by  3D or 1D electron--phonon (e-p) interaction in the Cu film (open circles and diamonds, respectively) using typical values $n=5$ ($n=3$ for 1D) and $\Sigma=2.1 \cdot 10^9$ WK$^{-5}$m$^{-3}$\cite{Sjltp}. In addition, for the 1D model we used $c_l=4900$ m/s for the longitudinal speed of sound in Cu. The only fitting parameter used was $\beta$, the fraction of dissipated heat flowing back to the normal metal, with a value $\beta=0.03$ kept constant for the three bath temperatures and for both models. As we see, the 3D e-p model reproduces the bulk data very well for all bath temperatures, but cannot explain the suspended sample data at all. We stress that since the electron gas volumes of the two samples are equal, there are no fitting parameters left, as we expect $\beta$ to be approximately the same for the bulk and suspended samples (both samples have the coolers on the bulk substrate). For the low temperature range, the suspended nanowire phonons are expected to be one-dimensional, thus we should also compare the suspended data with the one-dimensional e-p model \cite{Shekking} (diamonds, Fig. \ref{fig4}). It is clear that it cannot model the suspended data either (and adjusting $\beta$ does not improve the fit). The $T(V)$ cooling curves thus confirm the conclusion that e-p interaction is not the limiting dissipation mechanism in the suspended nanowires, unlike in the bulk samples.

\section{Details on SINIS thermometry}
In addition to using NIS junction as coolers, they were also used as sensitive thermometers  because of their highly non--linear and temperature dependent current--voltage (I--V) characteristics at sub-Kelvin temperatures \cite{SJukkareview,SkarvonenPRL}. In practice, the measurement is usually performed by connecting two junctions in series (SINIS), running a constant current through the junctions and measuring the temperature dependent voltage response.  When constant current biased, the SINIS-thermometer voltage $V_{\mathrm{therm}}$ is a sensitive function of temperature only, and this dependence can be calculated from the BCS--theory once the tunneling resistance of the junctions $R_{T}$ and the superconducting gap $\Delta$ are known. Since those parameters were always determined in a separate measurement of the I--V characteristics of the junctions, no free parameters are left, and the measured SINIS voltage can be unambiguously converted to temperature. We always checked this to be true by a calibration measurement, where the SINIS temperature was compared with the temperature given by a calibrated RuO thermometer while the refrigerator temperature was varied. The thermometer junctions were current biased with a bias resistor $R=10$ G$\Omega$ to ensure proper current bias even in the subgap, where junction resistance was typically $\sim$10 M$\Omega$ at low temperatures.  In the experiment, two different constant bias current values ($I \sim$ 10 pA) and ($ I\sim$100 pA) were used, the low one for low-temperature regime and higher one for high-temperature regime.  Two values were used because the SINIS temperature-to-voltage responsivity $\mathrm{d}V/\mathrm{d}T$ is a strong function of both $I$ and $T$ in such a way that the low bias (high bias) value gives a better responsivity at $T < 0.4$ K ($T > 0.4$ K). In addition, the lower bias value was always chosen higher than the measured two-electron/lifetime-broadened excess sub-gap current, to guarantee good response and no dependence on the details of the excess current mechanisms. $V_{\mathrm{therm}}$ was measured with a high input impedance differential voltage preamplifier (Ithaco 1201). In addition, while measuing the I--V characteristics, current was measured with a current preamplifier (Ithaco 1211).   

\begin{figure*}[ht!]
\begin{minipage}[t]{17pc}
\includegraphics[height=6cm]{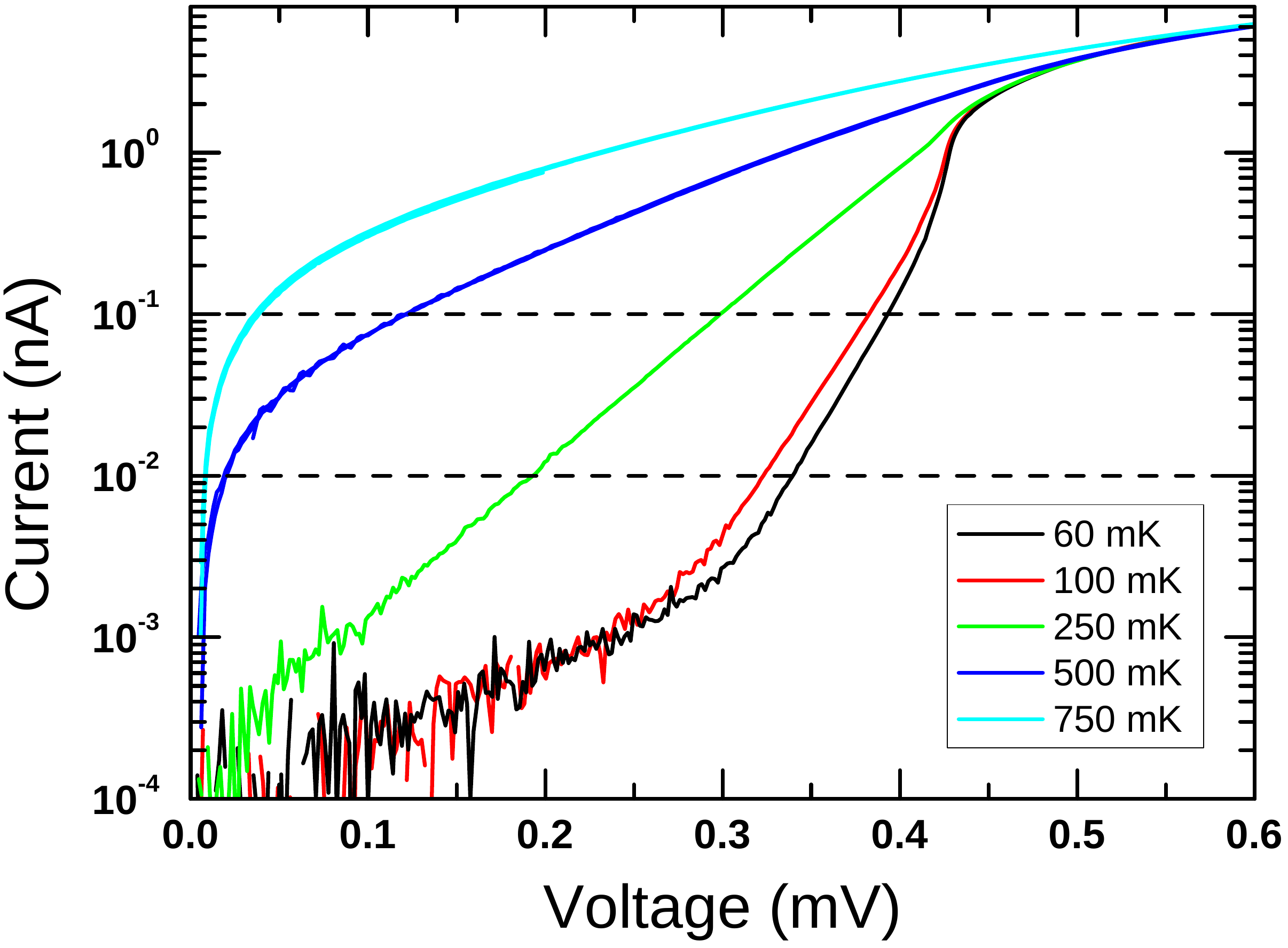}
\caption{\label{SF1}A measured sub--gap current--voltage characteristics of a typical SINIS thermometer ($2R_T=63$ k$\Omega$) at different bath temperatures in log-linear scale. Solid horizontal lines from top to down correspond to typical bias currents 100 pA and 10 pA, respectively.}
\end{minipage}\hspace{5pc}
\begin{minipage}[t]{20pc}
\includegraphics[height=6cm]{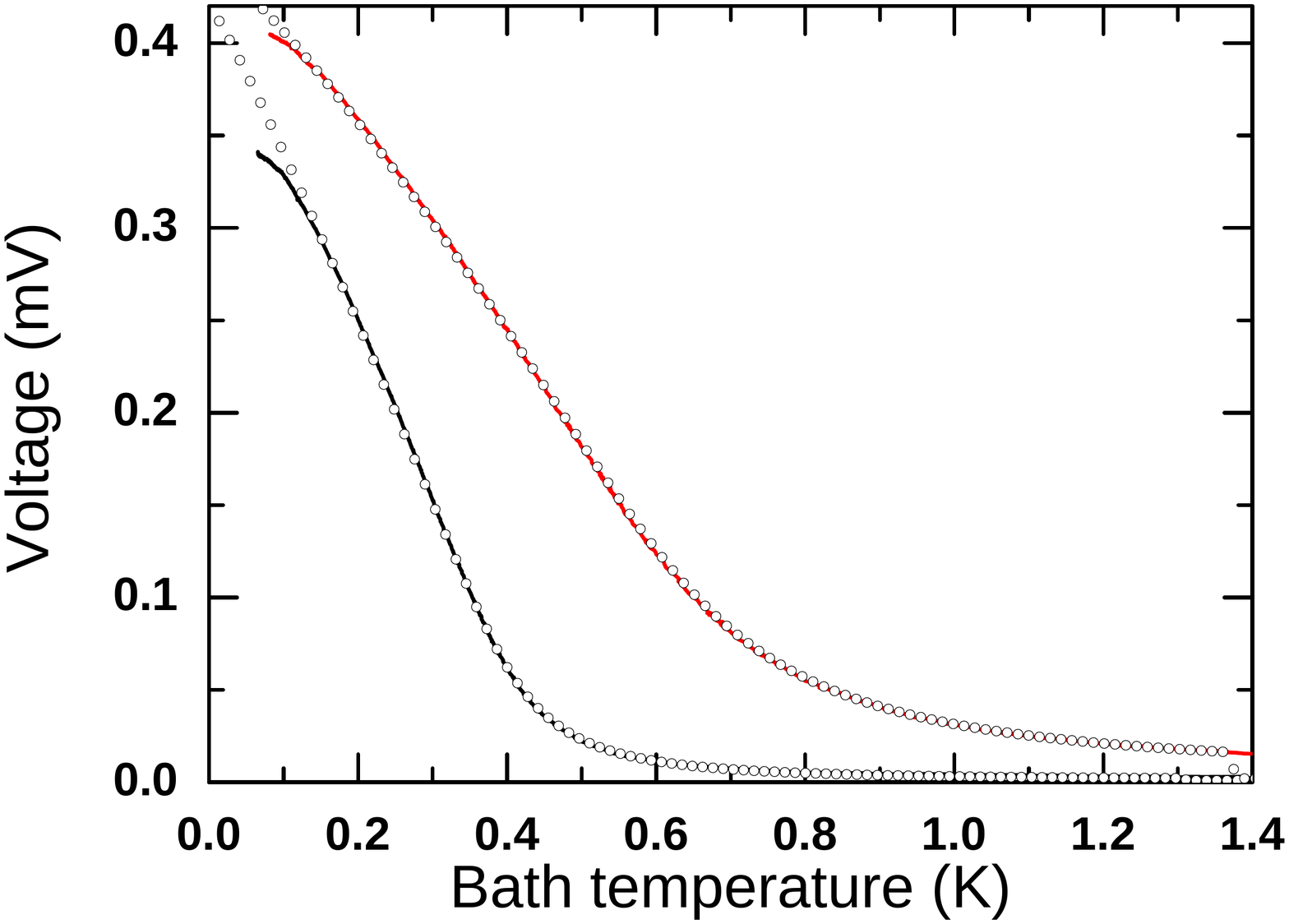}
\caption{\label{SF2}Measured voltage-bath temperature response of the SINIS thermometer in Fig. \ref{SF1} with two different 
values of bias current, 10 pA (black line) and 100 pA (red line) (same values as the lines in Fig. \ref{SF1}). Open circles were calculated using the single-particle tunneling Hamiltonian, Eq. \ref{IV}. Bath temperature was measured with a calibrated RuO thermometer.} 
\end{minipage} \end{figure*}

It is clearly seen that by choosing the bias current appropriately (two examples shown as horizontal lines in Fig. \ref{SF1}),  the responsivity $\mathrm{d}V/\mathrm{d}T$ of the thermometer can be adjusted to be best suited for the required temperature range. Fig. \ref{SF2} shows an example of the bath--temperature-to-voltage response (the bath temperature was measured with a calibrated RuO thermometer) of the SINIS thermometer in Fig. \ref{SF1} with bias current values 10 pA and 100 pA. With the low bias--current ($\sim$ 10 pA) there is a gain in the responsivity at low temperatures, while with the higher bias--current ($\sim$ 100 pA) the responsivity is better at higher temperatures $T > 0.4$ K. Best results with thermometry are obtained by repeating experiments with few different bias points for different temperature ranges. Note from Fig. \ref{SF1} that at current levels $I \sim 1$ pA temperature sensitivity is lost.

\subsection{IV characteristics and subgap conductance of NIS junctions}

The I--V characteristics can be understood by a simple BCS-based tunneling Hamiltonian theory \cite{Stinkham,SJukkareview}, which 
predicts  for a pair of identical normal metal-superconductor tunnel junctions (SINIS) 
\begin{equation}
\label{IV}
I(V)=\frac{1}{2eR_{T}}\int^{\infty}_{-\infty}g_{S}(E)[f_{N}(E-eV/2)-f_{N}(E+eV/2)]dE, 
\end{equation} 

where $R_T$ is the tunneling resistance of a single junction, $f_{N}(E)$ is the Fermi-function of the normal metal in 
quasiequilibrium and  $g_{S}(E)$ is the quasiparticle density of states (DOS) in the superconductor.
Note that equation (\ref{IV}) contains only the temperature of the normal metal and not that of the superconductor, and is valid as long as the quasiparticle distribution in the superconductor is also in quasiequilibrium (but not necessarily the same 
temperature). $g_{S}(E)$ can be determined from the BCS theory, which predicts in the weak coupling limit that 
$g_{S}(E)=|E|/\sqrt{E^2-\Delta^2}$, when $|E|>\Delta$ and $g_{S}(E)=0$ when $|E|<\Delta$ \cite{Stinkham}. However, in real 
materials there are processes that create quasiparticle states also within the gap $|E|<\Delta$. The easiest and most 
straightforward way to model these is with the so called Dynes-parameter, which was initially realized to model life--time 
broadening due to inelastic scattering (electron-electron, electron-phonon) \cite{SDynes}. This leads into a DOS of the form
\begin{equation}\label{DOS}
g_{S}(E)=\left|Re \left\{ \frac{E+i\Gamma}{\sqrt{(E+i\Gamma)^2-\Delta^2}} \right\}\right|, 
\end{equation}
where parameter $\Gamma$ describes the finite life--time ($\Gamma=\hbar/\tau$) of quasiparticle states in the superconductor. 

\begin{figure}
\includegraphics[width=8.5cm]{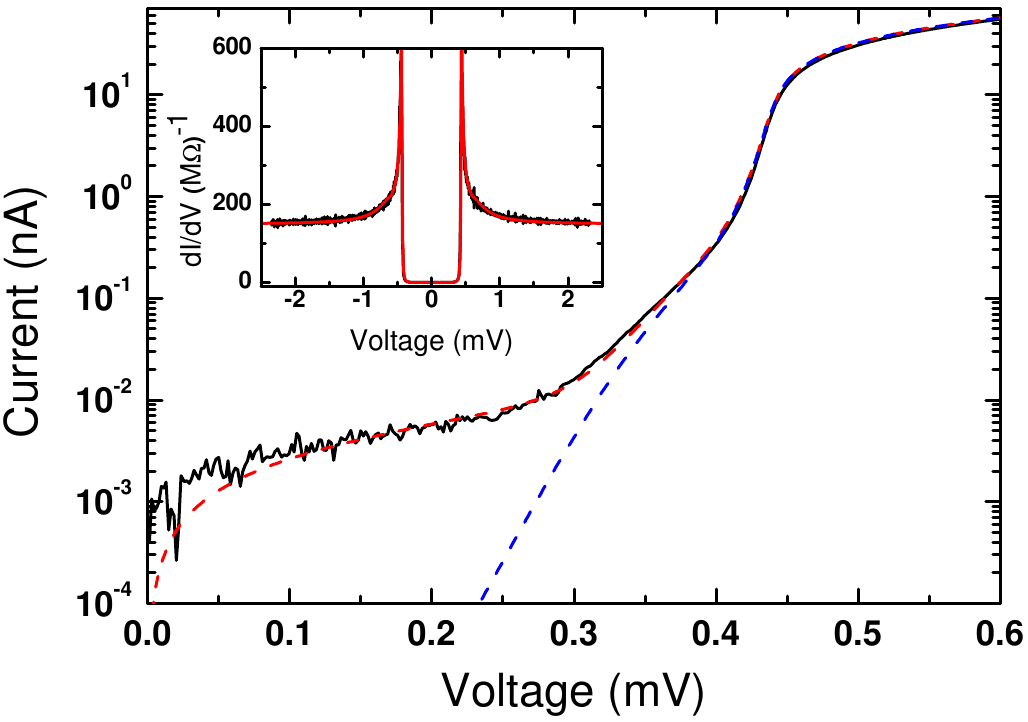}
\caption{\label{fig3} Sub--gap current voltage characteristics of SINIS Cooler junctions at $T_{\mathrm{bath}}=60$ mK. Experimental data is presented by a solid black line. Dashed lines show numerical calculations with broadened DOS $\Gamma/\Delta=2 \cdot 10^{-4}$ (red) and the case where life--time broadening is neglected (green). Inset shows the numerically calculated differential conductance $\mathrm{d}I/\mathrm{d}V$ of the measured data (black) and theory with $\Gamma/\Delta=2 \cdot 10^{-4}$ (red). $R_T=3.3$k$\Omega$.}
\end{figure}

Figure \ref{fig3} shows a typical measured suspended SINIS cooler I--V curve (black, solid line) in a logarithmic scale to 
highlight the sub--gap regime. As the cooler junctions are larger and have thus lower $R_T$, the current is higher than in the 
thermometer junctions (Fig. \ref{SF1}). This means that the sub-gap current is more easily measurable. The dashed, red line  in 
Fig. \ref{fig3} is a theoretical curve based on Eq. \ref{IV} that takes into account the broadened DOS by the Dynes model (dashed, red line), providing a good fit to the data with $\Gamma/\Delta=2\cdot 10^{-4}$ as the only fitting parameter. This value is consistent with what has been observed before for thin aluminum films \cite{SDynes, Sflyktman, Soneill}. On the contrary, without the broadening (dashed, blue line) the sub--gap I--Vs can not be fitted. Cooling of the junctions (apparent as dip around $\sim 0.4$ mV) is taken into account in the theoretical curves by using a voltage-dependent temperature that was obtained from the separate thermometer junctions. 

In addition to the finite life--time broadening, the microscopic nature of the finite sub--gap current can also originate from the higher order (Andreev) tunneling processes \cite{Shekking1,Shekkingleshouches}, which give a very similar looking result for the I-V curves \cite{Srajauria}. For the conclusions of this work it is irrelevant which microscopic sub-gap conduction process actually takes place, as both of them lead to anomalous heating at voltages below the gap and degradation of the cooler performance. However, with our observed small sub-gap current $\Gamma/\Delta=2\cdot 10^{-4}$ the anomalous heating is a small effect and just barely noticeable in the data at lowest temperatures (see Fig. 1(d) in the main text). 

\subsection{Temperature measurement in the low-temperature regime}

As was shown above, Eq. \ref{IV} gives an accurate description of the measured $I(V)$ curves, even at sub-gap voltages. This means that if we measure the $I(V)$ curve of the SINIS thermometer, we can determine $R_T$, $\Delta$ and $\Gamma$ unambiguously, and use Eq. (1) to convert the measured SINIS voltage to temperature without fitting parameters. We also point out that by biasing clearly above the sub-gap knee (Fig. \ref{SF1}) $\Gamma$ does not influence the results at all at the temperature range of interest, so that in practice only $R_T$ and $\Delta$ fix the temperature calibration. However, if one compares the measured responsivity curves $V_{\mathrm{SINIS}}$ vs. $T_{\mathrm{bath}}$  with the theory from Eq. \ref{IV} (Fig. \ref{SF2}), some deviation is clearly seen at the low temperature regime $T < 100 mK$, looking like a beginning saturation of $V_{\mathrm{SINIS}}$. This saturation is not an intrinsic limit for the junctions, since we have measured higher values of $V_{\mathrm{SINIS}}$ (lower temperatures) with the coolers operating. We conclude that the observed saturation is most likely caused by extrinsic heating power  radiated down the leads. This is plausible since a nanoscale sample can be overheated very easily due to the weakness of dissipation (electron-phonon, or phonon scattering). From the heating experiment (Fig. 3(a) in the main text) we can easily see that for our devices an excess power levels of $\sim 6$ fW  is enough to heat up the sample to 100 mK from $T_{\mathrm{bath}}=60$ mK. This kind of power levels are easily caused by Johnson noise power radiated down from the 4 K stage filters within the bandwidth of our low-temperature (below 4K) filters. However, the most important conclusion is that this observed saturation does not limit our temperature measurement for the coolers, as the theoretical self-calibration using Eq. \ref{IV} is still valid at $T < 100$ mK, regardless of the excess noise heating.       
       

\subsection{Lack of temperature gradients in the nanowire} 
\begin{figure}[ht!]
\includegraphics[width=8.5cm]{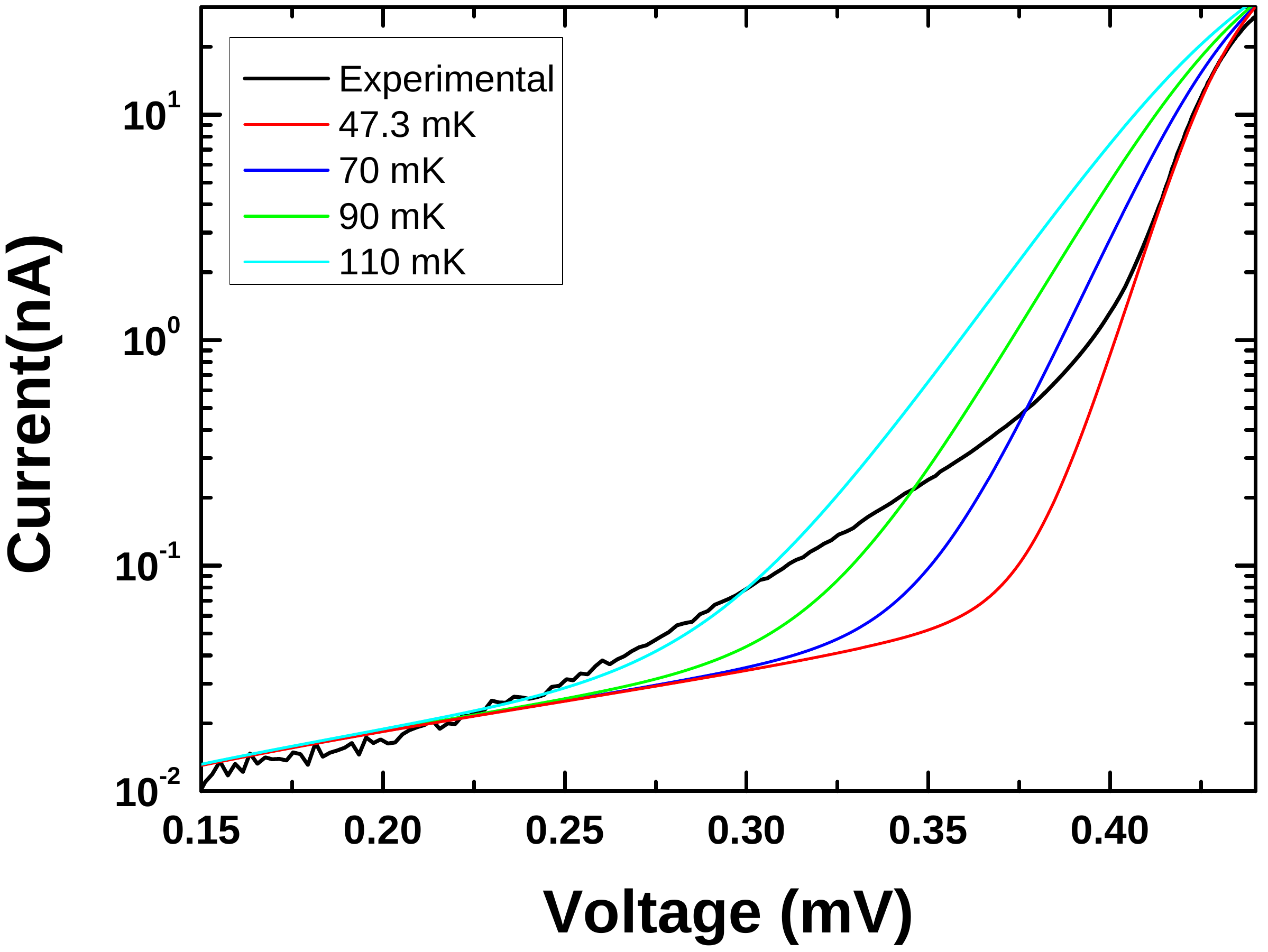}
\caption{\label{IVTe}The cooler $I$--$V$ curve of the sample (black line), whose $T(V)$ cooling curves were shown in Fig. 3, main text.  $T_{\mathrm{bath}}=60$ mK.  Theoretical curves (Eq. 1) with $T$ as a parameter are also shown, with $T=47.3$ mK giving the best fit at maximum cooling point around $V \sim 0.41$ V. }
\end{figure}

In the cooling experiments, we measured the temperature in the middle of the wire, whereas the coolers are actually located at the ends of the wire, a distance of $\sim$10--15 microns away (Fig. 1(c), main text). One might wonder whether any temperature gradients will develop within the nanowire, and whether the measured temperature is therefore unequal to the temperature at the cooler junctions. We investigated this question by comparing the measured $I$--$V$ characteristics of the cooler junctions with theoretical I-V curves with temperature as a parameter, Fig. \ref{IVTe}. It is clear that the cooler $I$--$V$ is consistent with lowest temperature $T = 47$ mK for this sample, which is the same temperature that was measured at optimal cooling for this sample at the SINIS thermometer in the middle of the nanowire (Fig. 4(b) main text). We conclude that no thermal gradients develop in the nanowire. This is consistent with the conclusion that phonon transmission at the nanowire--bulk boundary is the limiting dissipation mechanism.    

\subsection{Lack of influence of charging effects}
\begin{figure}[ht!]
\includegraphics[width=8.6cm]{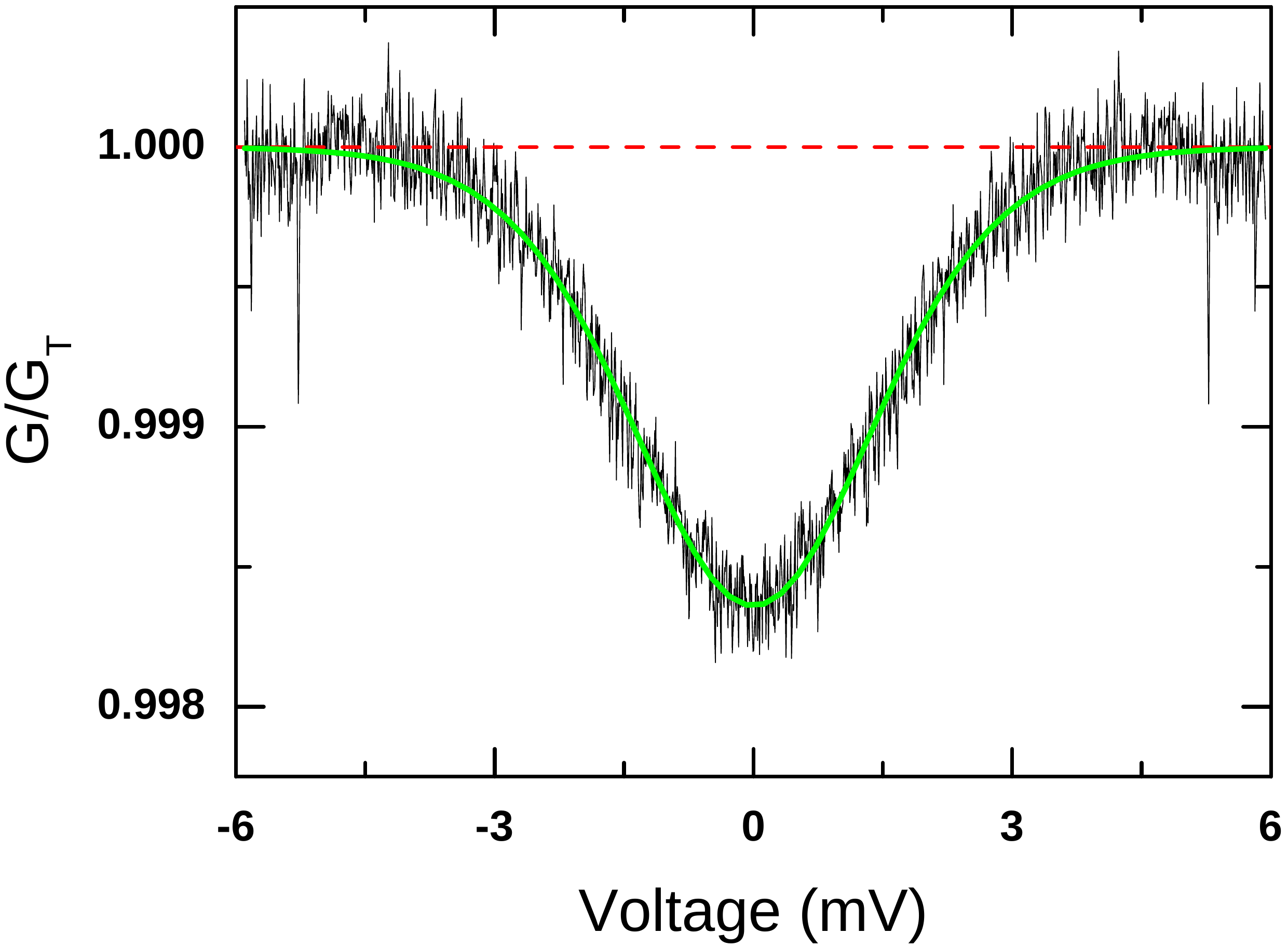}
\caption{\label{CBmeas}The measured differential conductance spectrum of suspended cooler at 4.2K. Conductance is normalized with $G_{T}$ (red, dashed line). The solid line is a one-parameter fit to  Eq. \ref{CB}, giving $E_C/k_B=20$ mK.}
\end{figure}

Coulomb blockade (charging effects) due to the small capacitance of sub-micron scale tunnel junctions can have a measurable effect on thermometry, especially in the limit $E_{C}>k_{B}T$ \cite{Scharging}, where $E_C=e^2/2C_{\Sigma}$ is the charging energy and $C_{\Sigma}=2C+C_{0}$ is the total capacitance of the junctions and island. 
 $E_C$ can be experimentally determined by measuring the tunneling conductance spectrum around zero bias in the weak Coulomb 
blockade limit $E_{C}<k_{B}T$, where a dip $\Delta G$ develops, depending on $E_C$ as \cite{SJukkareview}
\begin{equation}
\label{CB}
\frac{\Delta G}{G_{T}}=\frac{E_{C}}{3k_{B}T}, 
\end{equation}
where  $G_{T}$ is the tunneling conductance around $V=0$ without the dip. The measured charging energy at 4.2 K for a typical 
cooler sample is shown in Fig. \ref{CBmeas}. From the measurement we obtain $E_{C}/k_B=20$ mK, which is small enough not to affect the SINIS thermometry, i.e. analysis can be carried out with the simplest BCS--theory calculation, Eq. (1), without charging effects. This conclusion has been confirmed by a theoretical calculation of the SINIS response including charging effects \cite{Scharging}.


\begin{thebibliography}{28}
\bibitem[{not()}]{note}
\bibinfo{journal}{Note that for much shorter wires of length $\sim 1 \mu$m, 1D
  electron-phonon coupling will become dominant at low temperatures, so that
  phonon cooling will become impossible.}
\end{thebibliography}
\end{document}